\pageno=0
\nopagenumbers
\footline= { \ifnum\pageno>0 \hss\tenrm\folio\hss \fi}
\magnification=\magstep1 
\null
\def \e{{\rm e}}
\def \i{{\rm i}}
\def \m{\textstyle {1\over 2}}

\hsize=6.2truein
\vsize=8.5truein
\null \vskip 1in
\centerline{ \bf MINIJETS AND THE TWO-BODY PARTON CORRELATION}
\vskip .25in
\centerline{G. Calucci and D. Treleani}
\vskip .05in
\centerline{\it Dipartimento di Fisica Teorica dell'Universit\`a and INFN.}
\centerline{\it Trieste, I 34014 Italy}
\vfill
\centerline{ABSTRACT}
\vskip .25in
\midinsert
\narrower\narrower 
\noindent
A large number of double parton scatterings have been recently measured 
by CDF. The double parton scattering
process measures $\sigma_{eff}$, a non perturbative 
quantity related to the hadronic transverse size and with the dimensions of a 
cross section. The actual value
measured by CDF is considerably smaller as compared with the naive
expectation, namely the value of the 
inelastic non diffractive cross section. The 
small value of $\sigma_{eff}$ may be an effect of the hadron structure
in transverse plane. We discuss the problem  
by taking into account, at all orders, the two-body parton correlations in the
many-body parton distributions.
\endinsert 
\vfill
\eject
\par	{\bf 1. Introduction} 
\vskip 1pc
CDF has recently shown evidence of a large number of double parton 
collisions[1], namely events where, in the same inelastic interaction, 
two different pairs of partons scatter independently with large
momentum transfer.
The inclusive cross section for a double parton scattering 
is written as[2]:

$$\sigma_D=\int_{p_t^c}D_2(x_A,x_A';{\bf b}) \hat{\sigma}(x_A,x_B)
\hat{\sigma}(x_A',x_B') D_2(x_B,x_B';{\bf b})d{\bf b}dx_Adx_Bdx_A'dx_B'\eqno(1.1)$$
 
\par\noindent
$ \hat{\sigma}(x_A,x_B)$ is the 
parton-parton cross section integrated with the cut off $p_t^c$,
which is the lower 
threshold to observe final state partons as minijets, 
$x$ is the momentum fraction, $A$ and $B$ are labels to identify the two interacting 
hadrons. 
$\sigma_D$ is a function of the product $\hat{\sigma}(x_A,x_B)
\hat{\sigma}(x_A',x_B')$. Actually the two different partonic interactions are 
localized in two regions in transverse space with a size of order $(1/p_t^c)^2$
and at a relative distance of the order of the hadronic radius $R$,
in such a way that the two
partonic interactions add incoherently in the double scattering
cross section. The non perturbative input in Eq.(1.1)
is the two-body parton distribution $D_2(x,x';{\bf b})$, which
depends on the fractional momenta of the two partons taking part to the 
interaction and on their
relative transverse distance ${\bf b}$. The transverse distance
${\bf b}$ has to be the same for
the two partons of hadron $A$ and the two partons of hadron $B$, in order to have
the alinement which is needed for a
double collision to occur. $D_2$ is a dimensional quantity and therefore the process introduces a
non perturbative scale factor which is related to the hadronic transverse size. 
\par\noindent
The simplest possibility to consider is the one where the dependence of $D_2$ on 
the different variables is factorized:

$$D_2(x,x';{\bf b})=f_{eff}(x)f_{eff}(x')F({\bf b})\eqno(1.2)$$

\par\noindent
$f_{eff}$ is the effective parton distribution, namely the
gluon plus $4/9$ of the quark and anti-quark distributions
and $F({\bf b})$ is normalized to one. Multiparton
distribution are then uncorrelated and $D_2$ 
does not contain further informations
with respect to the one-body parton distribution (actually $f_{eff}$) 
apart form the dependence on ${\bf b}$, whose origin is the dimensionality of 
$D_2$ and which
gives rise to the scale factor $\sigma_{eff}$.
In fact in this case one may write

$$\sigma_D={\sigma_S^2\over\sigma_{eff}}\eqno(1.3)$$

\par\noindent
with

$${1\over\sigma_{eff}}=\int F^2(b)d^2b\eqno(1.4)$$

\par\noindent
and

$$\sigma_S=\int_{p_t^c}f_{eff}(x_A)f_{eff}(x_B)
  \hat{\sigma}(x_A,x_B)dx_Adx_B,\eqno(1.5)$$

\par\noindent
the single scattering expression of the perturbative QCD
parton model.
\par
Eq.(1.2) is the basic hypothesis
underlying the signature of a double parton collision which one has been
looking for in the experimental search[1,3]. 
The expected characteristic feature of a double collision 
is in fact that it should produce a final state analogous 
to the one obtained by super-posing two single scattering processes.
CDF measures:

$$\sigma_{eff}=14.5\pm1.7^{+1.7}_{-2.3}mb$$

\par	By looking at the dependence of $\sigma_{eff}$ on $x$ 
CDF has been able to verify the correctness of  
the factorization hypothesis in Eq.(1.2). The range of values of $x$ available
is limited to $x\le.2$, for the interaction producing a pair of minijets, and 
to $x\le.4$ for the 
interaction giving rise to a minijet and a photon.  
In the limited range of values of $x$ available,
the factorization hypothesis has shown to be consistent with the experimental evidence. 
\par	Since the uncorrelation hypothesis, as expressed in Eq.(1.2), 
is not inconsistent with experiment,
one can work out the case where all multiparton distributions are 
uncorrelated and one may look
for the sum of all multiparton interactions to the hadronic
inelastic cross section. The subset where all multiple parton collisions are
disconnected can be easily summed up in the uncorrelated case[4]. 
The result is
the semi-hard hadronic cross section $\sigma_H$, which represents the 
contribution to the hadronic inelastic cross section from events with
at least one semi-hard partonic interaction. The
actual expression is

$$\sigma_H=\int d^2\beta\Bigl[1-e^{-\sigma_SF(\beta)}\Bigr]
 =\sum_{n=1}^{\infty}\int d^2\beta{\bigl(\sigma_SF(\beta)\bigr)^n\over n!}
 e^{-\sigma_SF(\beta)}\eqno(1.6)$$

\par\noindent
The integration on the impact parameter of the 
hadronic collision, $\beta$, gives the dimensionality to the cross section. The
argument of the integral has the meaning of a Poissonian distribution
of multiple semi-hard partonic interactions with average number depending
on the impact parameter. 
\par	$\sigma_{eff}$ is related to $\sigma_H$ through Eq.(1.6). Indeed
the precise relation requires the knowledge of the functional dependence
of $F(\beta)$. Still the expression in Eq.(1.6) shows that $\sigma_H$ does
not depend on many details. The only relevant quantity is the value
of $\beta$ which makes $F(\beta)$ small, in such a way that the overall
argument of the integral becomes close to zero. In the simplest case 
where $F(\beta)={\rm exp}(-\beta^2/R^2)/\pi R^2$ one obtains a closed
analytic expression for $\sigma_H$. Actually

$$\sigma_H=2\pi R^2\bigl[\gamma+{\rm ln}\kappa+E_1(\kappa)\bigr]\eqno(1.7)$$

\par\noindent
where $\gamma=0.5772\dots$ is Euler's constant, $\kappa=\sigma_S/(2\pi R^2)$
and $E_1(x)$ is the exponential integral. For small $\kappa$ one obtains
$\sigma_H\to 2\pi R^2\kappa=\sigma_S$, for large $\kappa$, namely $\sigma_S\to
\infty$, one obtains $\sigma_H\to2\pi R^2\bigl(\gamma+{\rm}ln\kappa\big)$.
In this simplest example $\sigma_{eff}=2\pi R^2$. The value of $\sigma_H$
is therefore proportional to the measured value of $\sigma_{eff}$, the 
proportionality factor is slightly dependent on energy and on the cutoff. 
Sensible values of 
the hadron-hadron c.m. energy and of the cutoff give 
values for $\sigma_H$ which are some $30-40\%$ larger with
respect to the value of $\sigma_{eff}$. Although $\sigma_H$ has no
reason in principle to be close to $\sigma_{inel}$, a value of
$\sigma_H$ as small as $\sigma_{inel}/2$, which would result from
such considerations, seems nevertheless too small. Indeed $\sigma_H$,
as given in Eq.(1.6), has a smooth dependence on the infrared cutoff. If one
could really have the interaction under control also at small values
of $p_t^c$ one would say that $\sigma_H$ has to be
the same as the inelastic non diffractive cross section $\sigma_{inel}$. 
It would therefore be rather natural to expect for $\sigma_H$ 
a value not much smaller as compared to the value of 
$\sigma_{inel}$.
The large difference between $\sigma_H$ and $\sigma_{inel}$
could therefore be an indication that the simplest assumptions
underlying the derivation of the expression in Eq.(1.6) have to be revised.
\par	The main hypothesis which has been done to obtain
the expression for $\sigma_H$ in Eq.(1.6) is the Poissonian multiparton
distribution. On the other hand
one has to expect correlations between partons as a consequence
of the binding force. While most probably correlations will affect
the $x$ dependence of the multiparton distribution only for finite
values of $x$, and therefore at large rapidities, correlations
in the transverse parton coordinates are present in every kinematical
regime. Indeed the main reason of interest in multiple parton collisions, besides the
identification of the process itself, is precisely the measure of
the many-body parton correlations, which is an information on the hadron structure
independent on the one-body parton distributions
usually considered in hard processes. 
\par	In the present paper we work out the most general expression
for the semihard cross section $\sigma_H$, which one obtains by assuming
that only two-body parton correlations are present in the many-body parton
distributions and by summing all disconnected multiple parton
interactions. The two body parton correlation
in transverse plane is then worked out in great detail in a few cases
by considering different explicit shapes.
\vskip 2pc
\par	{\bf 2. General formalism and semi-hard cross section}
\vskip 1pc
\par	At a given resolution, provided by the cut off $p_t^{min}$
that defines the lower threshold for the production of minijets,
one can find the hadron in various partonic configurations. The probability of 
an exclusive $n$-parton distribution,
namely the probability to find the hadron in a configuration
with $n$-partons, is denoted by $W_n(u_1\dots u_n)$.
$u_i\equiv({\bf b}_i,x_i)$ represents the transverse partonic coordinate ${\bf b}_i$
and longitudinal fractional momentum $x_i$ while color and flavour variables
are not considered explicitly.
The distributions are symmetric in the variables $u_i$. One defines the generating 
functional of the multiparton distributions as:

$${\cal Z}[J]=\sum_n{1\over n!}\int J(u_1)\dots J(u_n)W_n(u_1\dots u_n)
du_1\dots du_n,\eqno(2.1)$$

\noindent
where the dependence on the infrared cutoff $p_t^{min}$ is implicitly understood,
and one may introduce also the logarithm of the generating 
functional: ${\cal F}[J]={\rm ln}\bigl({\cal Z}[J]\bigr)$.
The conservation of the probability yields the overall normalization condition

$${\cal Z}[1]=1.\eqno(2.2)$$

\noindent
 One may use the generating functional to derive the many body densities, 
i.e. the inclusive distributions
$D_n(u_1\dots u_n)$:

$$\eqalign{D_1(u)&=W_1(u)+\int W_2(u,u')du'+{1\over 2}
                    \int W_3(u,u',u'')du'du''+\dots\cr
                 &={\delta{\cal Z}\over \delta J(u)}\biggm|_{J=1}
                 ={\delta{\cal F}\over \delta J(u)}\biggm|_{J=1},\cr
     D_2(u_1,u_2)&=W_2(u_1,u_2)+\int W_3(u_1,u_2,u')du'+{1\over 2}
                    \int W_4(u_1,u_2,u',u'')du'du''\dots\cr
                 &={\delta^2{\cal Z}\over \delta J(u_1)\delta J(u_2)} 
                  \biggm|_{J=1}=
                   {\delta^2{\cal F}\over \delta J(u_1)\delta J(u_2)} 
                  \biggm|_{J=1}+
                  {\delta{\cal F}\over \delta J(u_1)}
                  {\delta{\cal F}\over \delta J(u_2)}  
                  \biggm|_{J=1}\cr
                 &\dots}\eqno(2.3)$$
 
\noindent
The 
many body parton correlations are defined 
by expanding ${\cal F}[J]$ in the vicinity of $J=1$: 

$$\eqalign{{\cal F}[J]=\int D(u)[J(u)-1]du+\sum_{n=2}^{\infty}{1\over n!}
\int C_n(&u_1\dots u_n)\bigl[J(u_1)-1\bigr]\dots\cr
  &\dots\bigl[J(u_n)-1\bigr]
du_1\dots du_n}\eqno(2.4).$$ 

\noindent
Here $D=D_1$ and the correlations $C_n$ describe how much the distribution
deviates from a Poisson distribution, which corresponds in fact to 
$C_n\equiv 0, n\ge 2$.
\par	By assuming the validity of the AGK cutting rules[5]
for semi-hard interactions one may express the semi-hard cross section
as the sum of all probabilities of multiple parton collisions: 

$$\sigma_H=\int d^2\beta\sigma_H(\beta)$$

with

$$\eqalign{\sigma_H(\beta)=\int&\sum_n{1\over n!}
  {\delta\over \delta J(u_1)}\dots 
  {\delta\over \delta J(u_n)}{\cal Z}_A[J]\cr
  \times&\sum_m{1\over m!}
  {\delta\over \delta J'(u_1'-\beta)}\dots 
  {\delta\over \delta J'(u_m'-\beta)}{\cal Z}_B[J']\cr
\times&\Bigl\{1-\prod_{i=1}^n\prod_{j=1}^m\bigl[1-\hat{\sigma}_{i,j}(u,u')\bigr]
   \Bigr\}\prod dudu'\Bigm|_{J=J'=0}} 
\eqno(2.5)$$

\noindent
where $\beta$ is the impact parameter between the two interacting hadrons
$A$ and $B$
and $\hat{\sigma}_{i,j}$ is the elementary probability for parton $i$ 
(of $A$) to have a hard interaction with parton $j$ (of $B$). 
The semi-hard cross section is constructed summing over all possible
partonic configurations of the two interacting hadrons (the sums over
$n$ and $m$) and, for each configuration with $n$ partons from $A$ and
$m$ partons from $B$, summing over all possible multiple partonic
interactions. This last sum is constructed asking for the 
probability of no interaction between the two configurations
( actually $\prod_{i=1}^n\prod_{j=1}^m[1-\hat{\sigma}_{i,j}]$ ). One
minus the probability of no interaction 
is equal to the sum over all 
semi-hard interaction probabilities.
\par	The presence of multiple parton interactions is
induced by the large flux of partons which is effective at large energies.
The most important contribution to the semi-hard 
cross section, as a consequence, is the contribution of
the disconnected partonic collisions,
namely the interactions where each parton undergoes at most
one semi-hard collision. These are, 
in fact, those multiple partonic interactions that, at a given number of
partonic collisions, maximize the parton flux. We simplify therefore the problem
by expanding the interaction probability
( the factor in curly brackets ) as sums and by removing all 
the addenda containing repeated indices:

$$\Bigl\{1-\prod_{i,j}^{n,m}\bigl[1-\hat{\sigma}_{ij}\bigr]\Bigr\}
\Rightarrow 
  \sum_{ij}\hat\sigma_{ij}-{1\over 2!}
  \sum_{ij}\sum_{k\not=i,l\not=j}\hat\sigma_{ij}\hat\sigma_{kl}+
  \dots\eqno(2.6)$$ 

\noindent
as a result the semi-hard cross section is constructed  with 
multiple disconnected 
parton collisions only, where disconnected refers to the perturbative
component of the interaction. Some features 
of semi-hard parton rescatterings are presented in Appendix A.
Because of the symmetry of the derivative operators in Eq.(2.5) one can 
replace the expression in Eq.(2.6) with:

$$nm\hat\sigma_{11}-{1\over 2!}n(n-1)m(m-1)\hat\sigma_{11}\hat\sigma_{22}
+\dots$$

\noindent
in such a way that the sums over $m$ and $n$ can be performed explicitly.
As a consequence the cross section at fixed impact parameter, 
$\sigma_H(\beta)$, can be expressed by the operatorial form:

$$\sigma_H(\beta)=
  \Bigl[ 1-\exp\bigl(-\delta\cdot\hat{\sigma}\cdot\delta'\bigr)\Bigr]
  {\cal Z}_A[J+1]{\cal Z}_B[J'+1]\Bigm|_{J=J'=0}
 \eqno(2.7)$$

\noindent
We have avoided writing explicitly the variables $u$ and $u'$ and
the functional derivative ${\delta /\delta J(u_i)}$
has been simply indicated as $\delta_i$.   
\par	The form of $\sigma_H(\beta)$ given by Eq.(2.6) is still too 
complicated to be worked out in its general form, since all possible 
multi-parton correlations are present in ${\cal Z}$. 
Therefore we further simplify the
problem by taking into account two-body parton correlations only.
Our explicit expression for ${\cal F}$ is therefore:

$${\cal F}_{A,B}[J+1]=\int D_{A,B}(u)J(u)du+
                    {1\over 2}\int C_{A,B}(u,v)J(u)J(v)dudv\eqno(2.8)$$

\noindent
where $D(u)$ is the average number of partons and 
$C(u,v)$ is the two-body parton correlation. 
\par
Either by using techniques of functional
integration or by means of a suitable diagrammatic expansion[6] 
one is able to obtain in this case a
closed expression for $\sigma_H(\beta)$:

$$\sigma_H(\beta)=1-\exp\Bigl[-
\m \sum _n a_n -\m \sum _n b_n /n\Bigr]
\eqno(2.9)$$
 
\noindent
where  $a_n$ and $b_n$ are functions of the impact parameter $\beta$ and are 
given by

 $$ \eqalign{a_n=\!\int D_A(u_1)\hat \sigma (u_1,u'_1) 
 C_B(u'_1-\beta,u'_2-\beta)\hat 
 \sigma (u'_2,u_2) C_A(u_2,u_3)&\dots\cr 
 \dots D_B(u'_n-\beta) 
 &\prod du_i du'_i }\eqno(2.10)$$

 $$ \eqalign{b_n=\!\int C_A(u_n ,u_1)\hat \sigma (u_1,u'_1) 
  C_B(u'_1-\beta,u'_2-\beta)\hat 
 \sigma (u'_2,u_2) &\dots\cr\dots  C_B(u'_{n-1}-\beta,u'_n-\beta)&\hat 
 \sigma (u'_n,u_n) \prod du_i du'_i\,.}\eqno(2.11) $$

The actual expression for $a_n$ (that will be referred as "open chain")  holds 
for $n$ odd. When $n$ is odd one may also have the symmetric case,
where the expression begins with $D_B$ and 
ends with $D_A$. When $n$ is even the initial and final 
distribution are either both $D_A$ or both $D_B$. In the definition of $b_n$,
that will be referred as "closed chain" $n$ is always even, so that one of
the ends is $A$ and the other is $B$. Sometimes the expression $n=2m$ will 
be used. One may notice
that, at a given order in the number of partonic interactions, one can obtain a 
term of kind {\it a} from a term
of kind {\it b} by replacing one $C$ with a pair of $D$'s. 
The operation can be done in $n$
ways. The combinatorial meaning of the $1/n$ factor multiplying 
each term of kind {\it b} in Eq.(2.9) is then 
understood. The factor $1/2$ in Eq.(2.9) is the consequence of the symmetry
between $A$ and $B$.
\par
 The cross section is given by an integral on the
impact parameter of the interaction probability, $\sigma_H(\beta)$, that is expressed as
one minus the probability of no interaction. The probability of no interaction
is given by the negative exponential of the sum over all possible different
connected structures, namely all structures of kind $a_n$ and of kind $b_n$.
With our approximations, Eq.(2.6) and Eq.(2.8), these are in fact all possible connected
structures which can be built with the average numbers $D_{A,B}$,
the two-body correlations $C_{A,B}$ and the interaction $\hat{\sigma}$ .
Expanding the exponential, the cross section can then be expressed as the 
sum over all possible structures, both connected and disconnected.
\par	One will notice that, when no correlations are present, all terms
of kind {\it b} disappear and only the first of the terms of kind {\it a},
namely $D_A\hat{\sigma}D_B$ is left. 
In that limit the cross section is given simply by:

$$\sigma_H=\int d^2\beta\biggl\{1-\exp\biggl[-
\int D_A(u-\beta)D_B(u')\hat{\sigma}(u,u')dudu'\biggr]\biggr\}
\eqno(2.12)$$

\par\noindent
which corresponds to the Poissonian distribution discussed in the
introduction.
\vskip 2pc

{\bf 3. An explicit case: Gaussian correlation }
\vskip 1pc
The bulk of the semi-hard cross section originates in the region of
small $x$ values. The experimental observation on the independence of
$\sigma_{eff}$ as a function of $x$, at small $x$-values, on the other hand
indicates that correlations in $x$, at small $x$, are not a strong
effect in the two-body parton distribution. A sensible case to consider is therefore
the one where there are no correlations
in fractional momenta, nor between 
fractional momenta and transverse coordinates, while only 
transverse coordinates are correlated:

$$D(u)=f(x)D({\bf b} )\qquad,\qquad C(u,u')=f(x) f(x') C({\bf b,b'}).
\eqno(3.1)$$

\noindent
Given the localization of the partonic collisions in transverse space,
the dependence of the parton-parton interaction probability on 
${\bf b}$ and ${\bf b}'$ is  represented as a $\delta$-function:

$$\hat \sigma (u,u')= \sigma_{x,x'} \delta ({\bf b-b'}).\eqno(3.2)$$

\noindent
All integrations on the fractional momenta $x$ and $x'$ are then factorized
from the integrations on the transverse coordinates and result in the single
scattering term cross section $\sigma_S$, whose expression is given in Eq.(1.5).
$a_n$ and $b_n$ are therefore considerably simplified: 

 $$ a_n=\sigma_S^n\int D_A({\bf b}_1) C_B({\bf b}_1-\beta,
  {\bf b}_2-\beta) C_A({\bf b}_2,{\bf b}_3)
\dots D_B({\bf b}_n-\beta) \prod d{\bf b}_i \eqno(3.3)$$

 $$ b_n=\sigma_S^n\int C_A({\bf b}_n ,{\bf b}_1) 
 C_B({\bf b}_1-\beta,{\bf b}_2-\beta) \dots 
 C_B({\bf b}_{n-1}-\beta,{\bf b}_n-\beta) \prod d{\bf b}_i\, ..\eqno(3.4)$$

\noindent
To proceed further one needs to consider explicit functional forms
for $C({\bf b}, {\bf b}')$.
A simplest extension of the Gaussian model discussed in the introduction
corresponds to the following Gaussian expression of the correlation term:

$$ D({\bf b })=h\exp [-F{\bf b }^2]\qquad,\qquad
   C({\bf b, b' })=k\exp [-\m F({\bf b+ b'})^2-\m G({\bf b- b'})^2].
\eqno(3.5)$$

\noindent
After defining ${\bf b}={\bf y}+\m \beta$, $b_n$ is explicitly written as: 

  $$b_n=\sigma_S^n k^n \int\exp [-\m nF\beta ^2] \exp [-(F+G) \sum {\bf y}_i^2]
    \exp[-(F-G) \sum  {\bf y}_i \cdot {\bf y}_{i+1}] \prod d{\bf y}_i,$$

\noindent
with the convention that the variable ${\bf y}_{n+1}$
coincides with ${\bf y}_1$.
\par\noindent
$b_n$ can then be worked out through the substitution

$$ {\bf v}_i =\mu {\bf y}_i -\nu {\bf y}_{i+1}\quad ;\quad
\mu -\nu =\sqrt {2F}\quad \mu +\nu =\sqrt {2G}\;,$$

\noindent
which leads to 

  $$b_n=J \sigma_S^n k^n \exp [-\m nF\beta ^2]\int
\exp[-\sum {\bf v} _i^2]\prod d{\bf v} _i \eqno (3.6)$$ 

\noindent
where the Jacobian, as discussed in Appendix B, is 
$ J= (\mu ^n -\nu ^n)^{-2}\,. $
$b_n$ is therefore finally written as:

$$ b_n=\sigma_S^n k^n \pi ^n (\mu ^n -\nu ^n)^{-2} \exp [-\m nF \beta ^2]\;.
\eqno(3.6') $$

\par
The case of the open chain, $a_n$, is less symmetric and requires a slightly
less straightforward treatment. Using the ${\bf y}$ variables, 
defined previously, the
expression of $a_n$ takes the form:

  $$a_n= \sigma_S^n h^2 k^{n-1}\exp [-\m nF\beta ^2] \int 
\exp [-(F+G)\,{\cal Y}^T \cdot {\cal M}\cdot  {\cal Y}] \prod d{\bf y}_i,
\eqno (3.7)$$

\noindent
where

$$ {\cal Y}^T=({\bf y_1,y_2,\dots,y_n})\eqno(3.8)$$

\noindent
and

$$ {\cal M}=\pmatrix{1+r&r&0&\ldots&0&0\cr
                 r&1&r&\ldots&0&0\cr
                 \vdots&\vdots&\vdots&\ddots&\vdots&\vdots\cr
                 0&0&0&\ldots&r&1+r\cr}\eqno(3.9) $$

$$r=\m (F-G)/(F+G)\,.\eqno(3.10)$$

\noindent
The Gaussian integral can be evaluated 
(details of the calculations are reported in Appendix B)
leading to the relation:

 $$a_n= \sigma_S^n h^2 k^{n-1} \pi ^n r^{n-1}[(2r+1)U_{n-1}(1/2r)]^{-1}
\exp [-\m nF\beta ^2]\;;\eqno(3.7')$$

\noindent
where $U_n$ is the Chebyshev polynomial of second kind[7].
\par
While the terms $a_n$ and $b_n$ are computed exactly, the sum of the series
in Eq.(2.9) can be performed only in limiting cases.
We discuss the case where the correlation length is much smaller 
with respect to the
hadronic radius. The parameter $G$ characterizes the correlation while
$F$ is related to the hadronic transverse size. Small correlation lengths correspond
to $F<<G$. We work out therefore the leading order term in $F$ 
while keeping the full structure in
$G$ both in $a_n$ and in $b_n$.
In order to find the limiting expression of $a_n$ when $F<<G$, one needs to work 
out the limit of $U_n(1/2r)$ for $1/2r \to -1+\epsilon$. One obtains:
 
$$U_n(-1+\epsilon) \approx (-)^n (n+1)$$

\noindent
By using this expression for $U_n$ and with similar, but simpler, manipulations 
one obtains the following limiting form for $a_n$:

$$a_n={Gh^2 \over 4nkF} Z^n\eqno(3.7'') $$

\noindent
with

$$ Z=\biggl({ 2\pi k\over G}\biggr)\sigma_S \exp [-\m F \beta^2] \eqno(3.11)$$

\noindent
The whole series of Eq.(3.1) can be then be summed yielding as a result:

$$S_a={Gh^2 \over 4kF}[-\ln (1-Z)].\eqno(3.12) $$

\par
To work out the limiting case for $b_n$ we keep the exact value of the difference
 $(\mu -\nu)^2=2F$, while setting $\mu \approx \nu \approx \sqrt {G/2}$
everywhere else. Keeping moreover into account 
the condition $n=2m$, we obtain:

$$b_{2m}={G \over 16Fm^2} Z^{2m} \eqno(3.6'')$$

\noindent
The resulting sum of the series in Eq.(3.1) is therefore given by

$$ S_b={G \over 32F} {\cal L}_3 (Z^2)\;.\eqno(3.13)$$

\noindent
where $ {\cal L}_3 (x)=\sum x^n /n^3 $ is the trilogarithm function[8].
\par
The final limiting form for the cross section, when $F<<G$, is therefore:

$$ \sigma_H=\int d^2\beta \sigma_H(\beta)=
\int d^2\beta \{1-\exp[-\m S_a -\m S_b]\}\,, \eqno (3.14)$$

\noindent
In the limit $G\to\infty$ the correlation goes to zero. In this case one
obtains $Z\to 0$, $S_a\to  (\pi\sigma_S/2kF) \exp [-\m F \beta^2] $ and
$S_b\to 0$. The expression of $\sigma_H$ in Eq.(3.14) is then reduced to
the Poissonian distribution discussed in the introduction.

\vskip 2pc
\par	{\bf 4. Two qualitatively different features of the correlation term}
\vskip 1pc
\par The uncorrelated multiparton distributions are characterized by two independent
features. One feature is that the $n$-body parton distribution is factorized in the product
of $n$ times the one-body parton distribution. The second
feature is that the distribution in
the number of partons (namely after integrating over all other degrees of
freedom) is a Poissonian. Both features are affected by the introduction of the two-body
correlations and it is therefore interesting to study the two effects separately.
\par
One may modify the number distribution, 
without introducing non-factorized
two-body correlations in ${\bf b}$, by using the factorized
expression 

$$C(u,u')=-\lambda D(u) D(u')\eqno(4.1)$$ 

\noindent
The terms appearing in 
Eq.s (2.10,11) are very much simplified in this case. After integrating over the 
longitudinal variables, the transverse integrations always appears
in the unique form

$$T\equiv T(\beta)=\sigma_S\int D_A({\bf b}) D_B({\bf b}-\beta) d{\bf b} \eqno (4.2)$$

\noindent
One therefore obtains:

$$\m \sum_n a_n= T/(1+ \lambda T)$$
$$\m \sum_m b_{2m}/2m=\ln(1- \lambda^2 T^2).$$

\noindent
and the cross section, at a fixed value of the impact parameter, is expressed as:

 $$ \sigma_H(\beta)=1-[1-\lambda^2  T^2]^{-1/2}
\exp [-T/(1+\lambda T)]\;.\eqno (4.3)$$

\noindent
The parameter $\lambda$ represents a measure of the deviation from the 
Poissonian distribution, both for the initial states and for 
the elementary collisions. In Appendix C some further elaboration on the factorized
case are presented. 

\par
Perhaps more interesting is the case where the number 
distribution of the incoming parton is Poissonian, because the correlation integrates
to zero, while the resulting distribution in the number of parton
collisions at fixed impact parameter is non-Poissonian, as
a consequence of the presence of the correlation term. 
To study this case a different way of dealing with the
approximation introduced in obtaining Eq.s (3.6'',7'') is useful.
\par
The alternative procedure is the following:
The variables ${\bf y}_i$ are substituted by the mean value 
$ {\bf Y}=\sum {\bf y}/n$ and by the differences 
${\bf x}_i={\bf y}_{i+1}-{\bf y}_i$, with the constraint 
$ \sum {\bf x}_i=0$. The relevant Jacobian is $J=1$ and so the integration 
volume is transformed according to:
$\prod d{\bf y}_i= d{\bf Y} \prod d{\bf x}_i \delta(\sum {\bf x}_i).$
Then in the terms containing $F$, which simply defines the size of 
the hadron, one
performs the substitutions ${\bf y}_i \approx {\bf Y}/n$ whereas the terms 
containing $G$ and the differences ${\bf x}_i$ 
are not modified. For the calculation of $a_n$, since only $n-1$ 
differences appear, the integrations in $d{\bf Y}$ and $d{\bf x}_i$ 
decouple in the Gaussian approximation and the result of Eq.s (3.7'',11) is
recovered; for the calculation of $b_n$ the constraint on the ${\bf x}_i$ is
essential, it may be implemented in the standard exponential form 
$\delta(\sum {\bf x}_i)=\int (2\pi )^{-2}\exp (\i\sum {\bf q \cdot x}_i) d{\bf q}$
and carrying out all the integrations the result of Eq. (3.6'') is reproduced.
\par
This alternative calculation suggests an initial distribution which can be easily 
worked out and which yields to a pure 
Poissonian distribution in the number of partons in the initial states. 
One may in fact consider the following expressions:

$$ D({\bf b })=h\exp [-F{\bf b }^2]\;,\;
  C({\bf b, b' })=-k' \Delta _2\exp [-\m F({\bf b+ b'})^2-\m G({\bf b- b'})^2]
\eqno(4.4)$$

\noindent
where $\Delta_2$ indicates the two-dimensional Laplace operator acting on
the transverse difference ${\bf b- b'}$ or ${\bf y- y'}$. 
By integrating $C$ over the
transverse variables one obtains zero. The number distribution depends therefore only
on $D$ and it is Poissonian by construction. The parameter $k'$ is
obviously related to the parameter $k$ in the previous paragraph.
$k$ was the strength of
the correlation at ${\bf b=b'}=0$. To keep the same normalization we make the
position $k'=k/G$.
\par
The contribution of all open chains $a_n$ with $n>1$ is therefore zero.
As previously observed in this case the differences have no constraint. Every independent integration over a transverse
difference corresponds therefore to an integration of a
single correlation term, which gives zero. Small nonzero effects have been 
possibly lost in the approximate substitution ${\bf y}_i \approx {\bf Y}/n$
in the terms containing $F$ in Eq.'s(3.6',7').
\par
When one computes the contribution of the closed chain $b_n$, the constraint is 
effective. The calculation is again performed by means of the exponential 
representation of the $\delta $-function and the resulting
expression can be brought to
the following form

$$ \tilde b_{2m}={G \over 8Fm^2} (Z/m)^{2m} (2m)! \eqno(4.5)$$

\noindent
The final result is obtained by summing all the terms above, with the
weight $(1/2m)$. The cross section (at $\beta$ fixed) is therefore

$$\sigma_H(\beta)=1-{\rm exp}\bigl[-T-\tilde S_b\bigr]\quad,\quad
   \tilde S_b=\sum_{m=1}^{\infty}
  {\tilde b_{2m}\over 2m}\eqno(4.6)$$

While a closed analytic form for the sum $\tilde S_b$ is not easily written in the
general case, in the limiting
case where the number of partonic collisions is large one may use
the Stirling approximation for the
factorial:
$m^{-m}\approx \sqrt {2 \pi m}\, \e^{-m} /m!$. With the help of the decomposition
$ (2m)!=2^{2m} m! \bigl(\m \bigr)_m$ the sum can be expressed as

$$\tilde S_b={\pi G \over 8F}\psi (w)\,,\eqno(4.7)$$ 

\noindent
where 

$$ w=(2Z/\e)^2\quad ,\quad
 \psi (w)=\sum _1 ^{\infty} { \bigl(\m \bigr)_m \over {m^2\,m!}}w^m$$

\noindent
$\psi(w)$ can be evaluated by computing, as intermediate step,

 $$(w\partial_w) (w\partial_w)\psi (w)=
\sum _1 ^{\infty} { \bigl(\m \bigr)_m \over m!}w^m\,,$$ 

\noindent
and by performing the integrations afterwards. The resulting expression is:

$$ \psi (w)=2{\cal L}_2 \bigl(\m[1-\sqrt{1-w}]\bigr)-
\ln^2 \bigl(\m[1+\sqrt{1+w}]\bigr)\;.\eqno(4.8)$$

\noindent
where 
 $ {\cal L}_2 (x)=\sum x^n /n^2 $ is the dilogarithm function.
\vskip 2pc
{\bf 5. Concluding discussion}
\vskip 1pc
The small value of $\sigma_{eff}$, the dimensional parameter
characterizing double parton scatterings, which has been measured
recently by CDF, is an indication that two-body parton correlations,
in the many-body parton distribution of the proton,
are likely to be sizeable.
In the case of an uncorrelated many-body parton distribution,
the value of $\sigma_{eff}$ puts a constraint on the
range of possible values of $\sigma_H$, the semi-hard contribution to the
hadronic inelastic cross section. The actual measured value of $\sigma_{eff}$
would give rise to values of $\sigma_H$ of the order of $\sigma_{inel}/2$
also at very large c.m. energies, where one would rather 
expect $\sigma_H\simeq\sigma_{inel}$.
The experimental evidence is also that, in the $x$ region accessible
experimentally namely at small $x$ values, the correlation in fractional
momenta is not a large effect. 
\par
In the present paper
we have worked out the semi-hard cross section $\sigma_H$ in the case where
initial state partons are correlated in the transverse parton coordinates.
In fact $\sigma_H$ can be worked out rather explicitly when only
two-body parton correlations are included
in the many-body parton distributions and when each parton can have at
most one semi-hard interaction. There are two qualitatively different features
in the two-body parton correlation, and both change the relation
between $\sigma_H$ and $\sigma_{eff}$ with respect to the uncorrelated case: 
\item{1-} The distribution in the
number of partons is not any more a Poissonian, although the dependence on the 
kinematical variables of the different partons is factorized.
\item{2-} The overall distribution in the number of partons, namely
after integrating on the partonic kinematical variables, is a Poissonian
but the dependence on the partonic kinematical variables is not factorized,
in which case the two-body parton correlation integrates to zero.
\par\noindent
The general case is obviously a combination of the two possibilities.
We point out however that
both cases separately can give rise to a small value of 
$\sigma_{eff}$ while keeping the 
value of $\sigma_H$ close to $\sigma_{inel}$.
\par
In the first case $\sigma_H$ is obtained by integrating the expression 
in Eq.(4.3). In the second by integrating the expression in Eq.(4.6).
The critical value of the impact parameter $\beta_c$, which gives the size
to the cross section $\sigma_H$, is the value which makes small the argument
of the exponential in the expression of $\sigma_H(\beta)$. The detailed 
dependence of the argument of the exponential at $\beta<\beta_c$ is not
of great importance for the determination of $\sigma_H$. It only determines
the degree of blackness, with respect to
semi-hard interactions, of the overlapping hadronic matter
at the given value of the impact parameter. 
It is therefore enough to determine the argument of the exponential
at the edge of the interaction region. In {\it case 1} 
the behaviour of the argument of the exponential when going
at the edge of the interaction region corresponds to the limiting behaviour
of the expression below at small $T$ values:

$${T\over 1+\lambda T}+{1\over2}{\rm ln}(1-\lambda^2T^2)
\to T=\sigma_S\int D_A({\bf b})D_B({\bf b}-\beta)d^2b$$

\noindent
while in {\it case 2} it corresponds to neglect 
$\tilde S_b$
as compared with $T$ in Eq.(4.6):

$$T+\tilde S_b
\to T=\sigma_S\int D_A({\bf b})D_B({\bf b}-\beta)d^2b.$$

\noindent
In both cases the
critical value $\beta_c$ has the same value which one
finds for the uncorrelated distribution and therefore $\sigma_H$,
as in the uncorrelated case, is roughly equal to $2\pi/F$, 
where $F$ is the parameter giving
the extension of the one-body parton distribution in the transverse plane.
\par
When the double scattering cross section is worked out by expanding 
the expression of $\sigma_H$, one finds
for the effective cross section the relation

$$\sigma_{eff}={2\pi\over F\cdot(1+\lambda)^2}$$

\noindent
in {\it case 1}, and 

$${1\over\sigma_{eff}}={F\over2\pi}+\int C({\bf b}, {\bf b}')
C({\bf b}-\beta, {\bf b}'-\beta)d^2bd^2b'd^2\beta$$ 

\noindent
in {\it case 2}.
\par
A qualitative feature is that in both cases one obtains a 
value of $\sigma_{eff}$ which may be sizably smaller
with respect to $2\pi/F\simeq\sigma_H$.
While, on the other hand, nothing prevents the value of $\sigma_H$ to
be close to the value of $\sigma_{inel}$. 
The smaller value of $\sigma_{eff}$, with respect to the 
expectation of the uncorrelated case, is
rather generally
associated with the increased dispersion of the distribution
in the number of partonic collisions: In the case of no correlations
the distribution is strictly Poissonian when the impact parameter is fixed.
When correlations are introduced the
distribution in the number of parton collisions, at fixed $\beta$,
is not Poissonian any more
and the natural consequence is that the dispersion in the number of collisions 
is increased.
\par
The indication from the measure of the rate of double parton scatterings
is therefore that two-body parton correlations are likely to be important
while, unfortunately, one cannot say much about dynamical quantities, like the 
the correlation
length. Useful observables to be measured, 
in order to get some more insight into the problem,
would be the semi-hard cross section $\sigma_H$ and the triple parton
scattering cross section. Our present analysis shows that
$\sigma_H$ can be reliably discussed in perturbation theory. 
The measure of $\sigma_H$, in association with $\sigma_{eff}$, would 
help considerably in clarifying the size of the effect induced by the presence
of the two-body parton correlations: All the considerations of the present
paper are based on the prejudice that $\sigma_H$ should have a value
rather close to the value of $\sigma_{inel}$. 
\par\noindent
The measure of triple parton scattering would allow one to test
the possibility discussed in {\it case 1}, since in that case the
rate of triple scatterings would be strictly fixed by the measured rate of
double scatterings. In the other cases presently discussed the knowledge of the
rate of triple parton collisions would allow one to obtain the actual
values of the parameters of the correlation term.
\par
A lot of effort has been put on the study of the proton structure
as a function of the momentum fraction $x$. The distribution of
partons however depends on three degrees of freedom, the momentum
fraction $x$  and the transverse parton coordinate ${\bf b}$.
The measure of the rate of triple 
and of higher order partonic collisions is the essential tool
to learn on the parton structure of the proton in transverse
plane.

\vskip 2pc
{\bf Acknowledgements}
\vskip 1pc
\par\noindent
We thank M. Strikman for useful comments and discussions.
\par\noindent
This work was partially supported by the Italian Ministry of University and of
Scientific and Technological Research by means of the Fondi per la Ricerca
scientifica - Universit\`a di Trieste.
\vfill
\eject 
{\bf Appendix A}
\vskip 1pc
\par
In this appendix some short observations about the effects of the rescattering
processes will be displayed. The general form of the hadron-hadron
scattering cross section Eq. (2.5) has been reduced to a more manageable form 
by throwing away all the rescattering process by means of the position:
$$\Bigl\{1-\prod_{i,j}^{n,m}\bigl[1-\hat{\sigma}_{ij}\bigr]\Bigr\}
\Rightarrow 
  \sum_{ij}\hat\sigma_{ij}-{1\over 2!}
  \sum_{ij}\sum_{k\not=i,l\not=j}\hat\sigma_{ij}\hat\sigma_{kl}+
  \dots\eqno(A.1)$$ 
It is possible to deal with some rescattering process without too much
effort if we look at processes where only one of
 the two colliding partons has already suffered some collision. In this case
instead of Eq. (A.1) we write 
$$\Bigl\{1-\prod_{i,j}^{n,m}\bigl[1-\hat{\sigma}_{ij}\bigr]\Bigr\}
\Rightarrow 
  \sum_{ij}\hat\sigma_{ij}-{1\over 2!} \sum_{ij}\Bigl\{
  \sum_{k\not=i,l}\hat\sigma_{ij}\hat\sigma_{kl}+
  \sum_{k,l\not=j}\hat\sigma_{ij}\hat\sigma_{kl}-
  \sum_{k\not=i,l\not=j}\hat\sigma_{ij}\hat\sigma_{kl}\Bigr\}+
  \dots\eqno(A.2)$$ 
In this expression the first term shows that the partons belonging to the first
hadron suffer only one collision, the partons belonging to the second hadron
undergo any number of collisions, the second term describes the symmetrical 
situation, the third term eliminates the double counting of the 
single-collision processes.
\footnote*{This possibility was already considered in ref [6], for an
unsymmetrical situation}.

The operatorial form of the cross section at fixed impact parameter 
$\sigma_H(\beta)$ acquires now the more complicated shape:
$$\eqalign{ \sigma_H(\beta)=
  \Bigl[ 1-& \exp\bigl(\delta\cdot[\e^{\hat{\sigma}\cdot\delta'}-1]\bigr)
          - \exp\bigl(\delta'\cdot[\e^{\hat{\sigma}\cdot\delta}-1]\bigr)
          + \exp\bigl(-\delta\cdot\hat{\sigma}\cdot\delta'\bigr)\Bigr]\times\cr
           & {\cal Z}_A[J+1]{\cal Z}_B[J'+1]\Bigm|_{J=J'=0}}
 \eqno(A.3)$$
This expression seems of uneasy interpretation but is gives some information
when reduced to simpler particular cases
The simplest, but non trivial result is produced when the incoming parton
distribution is purely Poissonian and only the double scattering is actually
considered. In these situations it is easily seen that instead of Eq. (2.10)
one obtains the form:
$$\eqalign{\sigma_H=&\int d\beta\biggl\{1-\exp\Bigl[-
\int D_A(u)\hat{\sigma}(u,u')D_B(u')dudu'\Bigr]\times\cr
 &\biggl(\exp\Bigl[\m \int D_A(u)\hat{\sigma}(u,u')\hat{\sigma}(u,u'')
D_B(u')D_B(u'')dudu'du''\Bigr]+\cr
 &\exp\Bigl[\m \int D_A(u) D_A(u'')\hat{\sigma}(u,u')\hat{\sigma}(u'',u')
D_B(u')dudu'du''\Bigr]-1\biggr)\biggr\}}
\eqno(A.4)$$
This form is not unexpected, it could have been written by hand, Eq.(A.3)
shows a more systematic way of deriving it and possible further corrections.
The final distribution of the collision is $not$ Poissonian and this deviation
is purely due to the hard dynamics. It can be observed that precisely this 
origin in the hard collision may also offer a way to distinguish these effects
from the effects of correlations in the incoming two-body distributions, in
fact in the case here sketched there should be an unbalance among the
kinematical variables of the pair of jets originating from the scattered
partons.
\vskip 2pc
{\bf Appendix B}
\vskip 1pc
The Jacobian appearing in Eq.(3.6), which
arises from the transformation from the variables {\bf y} to the
variables {\bf v}, is expressed as $J=[det {\cal J}]^{-2}$. The actual form of
  ${\cal J}$ is:
 $${\cal J}=\pmatrix{\mu&-\nu&0&\ldots&0&0\cr
                 0&\mu&-\nu&\ldots&0&0\cr
                 \vdots&\vdots&\vdots&\ddots&\vdots&\vdots\cr
                 -\nu&0&0&\ldots&0&\mu\cr} $$
and the result $det{\cal J}=\mu ^n-\nu ^n$ is easily obtained by expansion
according to the last row. The exponent $-2$ appears because the matrix 
describes the inverse transformation and there are two transverse dimensions.
\par
In order to calculate the determinant of the matrix $ {\cal M} $ 
in Eq.(3.7) an 
auxiliary matrix $ {\cal A} $ is introduced so that we have
$$ {\cal M}=\pmatrix{1+r&r&0&\ldots&0&0\cr
                 r&1&r&\ldots&0&0\cr
                 \vdots&\vdots&\vdots&\ddots&\vdots&\vdots\cr
                 0&0&0&\ldots&r&1+r\cr} \quad ,\quad
 {\cal A}=\pmatrix{1&r&0&\ldots&0&0\cr
                 r&1&r&\ldots&0&0\cr
                 \vdots&\vdots&\vdots&\ddots&\vdots&\vdots\cr
                 0&0&0&\ldots&r&1\cr} $$

By means of the standard rules for the computation of the determinants two
recurrence relations are obtained:
$$det {\cal M}_n=det {\cal A}_n +2r\,det {\cal A}_{n-1} +
r^2 det {\cal A}_{n_2}$$
$$det {\cal A}_n= det {\cal A}_{n-1} -r^2 det {\cal A}_{n_2}$$
which imply the simpler relation
$$det {\cal M}_n= (2r+1)det {\cal A}_{n-1}$$
The recurrence relations for the determinants of  ${\cal A}_n$ are very similar
to the recurrence relation for the Chebyshev polynomials $U_n$ [7] and in this
way it is possible to get the final expression for them
$$det {\cal A}_n=r^n U_n(1/2r)\,.$$
For completeness we give also the explicit expression of the 
$U$-polynomials:
$$U_n(\cos \theta)={\sin (n+1) \theta \over \sin \theta}\;.$$
\vskip 2pc
{\bf  Appendix C}
\vskip 1pc
In this appendix we shall briefly discuss some non-Poissonian one-body 
densities, the essential point were already shown in ref.[6], but it may be
useful to state again them in order to have a comparison of the previous 
treatment. The starting point of Eq.(2.7) is specialized to the case of 
pure one-body densities in the form
$$ \sigma _H =(1-\exp T\partial\partial')X_A({\cal Y})X_B({\cal Y'})
|_{{\cal Y}={\cal Y}_o,{\cal Y'}={\cal Y'}_o}.\eqno (C.1)$$
The Poissonian distribution is given by $X=\exp[{\cal Y}-{\cal Y}_o]$, the
situation described in Eq.(4.1) corresponds to $$X=\exp[{\cal Y}-{\cal Y}_o]
 +\m \lambda [{\cal Y}-{\cal Y}_o]^2\,,\eqno (C.2)$$
Then by applying Eq.(C.1) to Eq.(C.2), one gets as intermediate step
$$ \exp [-T\partial'+\m \lambda (T\partial')^2]\exp[{\cal Y}'-{\cal Y}'_o]
 +\m \lambda [{\cal Y}'-{\cal Y}'_o]^2 |_{{\cal Y'}={\cal Y'}_o} ,\eqno (C.3)$$
By the use of the general formula:
$$\exp[-p\partial_x^2]\exp[qx^2]=
{1\over {\sqrt {1-4pq}}}\exp\Bigl[{qx^2 \over {1+4pq}}\Bigr]$$
it is seen that the expression appearing in Eq.(C.3) reduces to the form
already given in Eq.(4.3). 
\par\noindent
One could also choose the form 
$$X=[1-({\cal Y}-{\cal Y}_o)]^{-\alpha}$$
which corresponds to a negative binomial distribution for the initial partons
and obtain in this way the result
$$\sigma_H(\beta)=1-T^{-\alpha}\,U[\alpha;1;1/T]\,,$$
by representing the incoming distributions as Laplace transforms as shown in 
ref.[6], note that a slight simplification has been introduced in the 
distribution, with respect to that reference, and this reflects into a slight 
simplification of the result, which keep however its main properties, in 
particular $U$ is the irregular confluent hypergeometric function.
\vfill
\eject
{\bf References}
\vskip 1pc
\item{1.} F. Abe et al. (CDF Collaboration), submitted to 
{\it Phys. Rev.} {\bf D} April 14, 1997;
FERMILAB-PUB-97/094-E.
\item{2.} C. Goebel, F. Halzen and D.M. Scott, {\it Phys. Rev.}
{\bf D22}, 2789 (1980);
N. Paver and D. Treleani, {\it Nuovo Cimento} {\bf A70},
215 (1982); B. Humpert, {\it Phys. Lett.} 
{\bf B131}, 461 (1983); B. Humpert and R. Odorico, {\it ibid} {\bf 154B}, 211 (1985); T. Sjostrand and
M. Van Zijl, {\it Phys. Rev.} {\bf D36}, 2019 (1987).
\item{3.} T. Akesson et al. (AFS Collaboration), {\it Z. Phys.} {\bf C34}
163 (1987); J. Alitti et al. (UA2 Collaboration), {\it Phys. Lett} {\bf B268}
145 (1991); F. Abe et al. (CDF Collaboration), {\it Phys. Rev.} {\bf D47}
4857 (1993).
\item{4.} Ll. Ametller and D. Treleani, {\it Int. J. Mod. Phys.} {\bf A3}, 521 
(1988). 
\item{5.} V. Abramovskii, V.N. Gribov and O.V. Kancheli, {\it Yad. Fiz.}
{\bf 18}, 595 (1973) [{\it Sov. J. Nucl. Phys.} {\bf 18}, 308 (1974) ];
I.G. Halliday and C.T. Sacharajda, {\it Phys. Rev.} {\bf D8}, 3598 (1973);
J. Koplik and A.H. Mueller, {\it Phys. Lett.} {\bf 58B}, 166 (1975);
L.D. Mc Lerran and J.H. Weiss, {\it Nucl. Phys.} {\bf B100}, 329 (1975);
L. Bertocchi and D. Treleani, {\it J. Phys.} {\bf G3} 147 (1977);
V.M. Braun and Yu.M. Shabelski, {\it Int. J. Mod. Phys.}
{\bf A3}, 2417 (1988);
G. Calucci and D. Treleani {\it Phys. Rev.} {\bf D49}, 138 (1994);
{\bf D50}, 4703 (1994);
J. Bartels and M. W\"{u}sthoff, {\it Z. Phys.} {\bf C66}, 157 
(1995); D. Treleani, {\it Int. J. Mod. Phys.} {\bf A11} 613 (1996).
\item{6.} G. Calucci and D. Treleani, {\it Nucl. Phys.} 
{\bf B} (Proc. Suppl.) 18C, 187 (1990) and 
{\it Int. J. Mod. Phys.} {\bf A6}, 4375 (1991).
\item{7.} M. Abramowitz and I. A. Stegun, {\it Handbook of Mathematical
Functions}, Dover Publications, Inc., New York.
\item{8.} L. Lewin, {\it Dilogarithms and Associated Functions},
Macdonald, London (1985).
\vfill   
\end
\bye